\documentclass{iopart}
\usepackage{iopams}
\usepackage{epsf}
\newcommand{\mathbold}[1]{\mbox{\boldmath $#1$}}

\begin{document}
\title[Black Hole Binaries in Globular Clusters]{Gravitational Radiation from Black Hole Binaries in Globular Clusters}
\author{M J Benacquista\dag}
\address{\dag\ Dept. of Sciences, Montana State University-Billings, Billings, Montana, 59101}
\ead{benacquista@msubillings.edu}

\begin{abstract}
A populations of stellar mass black hole binaries may exist in globular clusters. The dynamics of globular cluster evolution imply that there may be at most one black hole binary is a globular cluster. The population of binaries are expected to have orbital periods greater than a few hours and to have a thermal distribution of eccentricities. In the LISA band, the gravitational wave signal from these binaries will consist of several of the higher harmonics of the orbital frequency. A Monte Carlo simulation of the galactic globular cluster system indicates that LISA will detect binaries in 10\% of the clusters with an angular resolution sufficient to identify the host cluster of the binary.
\end{abstract}

\submitto{\CQG}
\maketitle

\section{Introduction}

Globular clusters may be the first isolated stellar systems to have their population of relativistic binaries studied with gravitational radiation. The galactic globular cluster system consists of about 150 clusters scattered throughout the halo of the galaxy~\cite{harris96}. At an age of nearly 13 Gyr, they consist of old population II stars and stellar remnants. The dynamical evolution of a globular cluster will tend to concentrate the more massive stars and binaries into the core of the cluster. This will enhance the likelihood of encounters between binaries and single stars, which tends to increase the number of relativistic binaries in globular clusters. These binaries will be potential sources for LISA~\cite{benacquista01}.

Although there is no direct observational evidence for black hole binaries in globular clusters, there are two theoretical scenarios which are expected to produce black holes in globular clusters. Miller and Hamilton~\cite{miller01} describe a mechanism whereby intermediate mass black holes will be produced in globular clusters with high central densities. These black holes will have masses $\sim 1000~M_{\odot}$ and may reside in the cores of up to 20 globular clusters. As compact objects in the core of these globular clusters pass close to the intermediate mass black hole, they may emit gravitational radiation within the frequency band of LISA. This radiation will be in the form of bursts. There is currently evidence of an intermediate mass black hole in M15~\cite{gebhardt00}.

The scenario of Portegies Zwart and McMillan~\cite{portegieszwart00} allows for the production of numerous stellar mass black hole binaries in the cores of globular clusters. These are then ejected from the cluster through interactions with other black holes and binaries in the cluster. The last black hole binaries in the cluster will have no other black holes to eject it. Thus, the last dynamical encounter can have three possible outcomes: the binary and the encountering black hole can all be ejected, the binary can be ejected and the encountering black hole can be retained, the binary can be retained and the encountering black hole can be ejected. Although these three outcomes are not equally likely, the probability of retention of a black hole binary can be estimated to be between 30\% and 50\%. Consequently, between 30\% and 50\% of globular clusters can be expected to have one remaining stellar mass black hole binary. This highly eccentric binaries of this population are viable sources for LISA.

\section{Stellar Mass Black Holes}

Stellar mass black holes with $M \sim 10~M_{\odot}$ are formed in globular clusters within the first $10^8~{\rm yr}$ of its history. These objects will segregate to the core of the cluster within the first few relaxation times. Barring any massive seed black holes, these objects will be the most massive objects in the core. Consequently, they will effectively form their own cluster and begin to interact solely with themselves in the presence of the background potential of the cluster. Encounters with the other stars in the cluster will provide little more than slight perturbations to the subsequent dynamical evolution of the black holes. This evolution will lead to the ejection of nearly all of the black holes in the cluster. The ejected black holes were studied as possible sources for LIGO by Portegies Zwart and McMillan~\cite{portegieszwart00}. However, the dynamical evolution may also leave a few longer period black hole binaries in the cores of globular clusters. These are the potential sources for LISA.

The process by which dynamical evolution leads to the ejection of black holes from globular clusters also leads to the hardening or shrinking of the binary systems. The outcome of a gravitational encounter between a binary and a field star in a cluster depends upon the binding energy of the binary and the masses of all three bodies involved in the encounter. If the binding energy of the binary is greater than the kinetic energy of the field star, then the common result is a transfer of binding energy to the kinetic energy of the three stars involved in the encounter. The least massive of the three stars is ejected from the encounter and the remaining two form a tighter binary. The binding energy of the system generally increases by about 20\% in an encounter and roughly 1/3 of the liberated gravitational energy goes into recoil energy of the binary. The process results in the gradual ejection of both binary and single black holes from the globular cluster as they gain sufficient velocity to escape the potential of the cluster.

This process is expected to take roughly 2 Gyrs. At the end of this time, the most likely result is three black holes remaining in the form of a binary and a single black hole. The last binary/single interaction can result in three possible outcomes: all three black holes can receive sufficient recoil to escape from the cluster, a binary can remain and eject a single black hole, or the single black hole can remain and eject the binary. The last outcome is highly unlikely as it would require the single black hole to be more massive than the binary. The more likely outcome would then be the exchange of the massive black hole with the least massive of the original binary components, resulting in the ejection of the least massive black hole and retention of the newly created massive binary. Thus, we can expect that roughly half of the globular clusters to contain a retained black hole binary.

The binaries which remain in the globular clusters will have a thermal distribution of eccentricities so that $P(e) = 2e$. We can determine the minimum orbital period expected for a retained binary by looking at the maximum orbital period of an ejected binary. Following Portegies Zwart and McMillan, a binary will be ejected if
\begin{equation}
E_{\rm b} \gtrsim 36 W_0\frac{m_{\rm bh}}{\langle{m}\rangle} kT
\label{minimum_binding}
\end{equation}
where $W_0 = \langle{m_{\rm bh}}\rangle \vert{\phi_0}\vert/ kT$ is the dimensionless central potential of the cluster and $(3/2) kT$ is the average kinetic energy of a cluster star. The average cluster star mass is $\langle{m}\rangle$ and $m_{\rm bh}$ is the mass of each black hole in the binary. If the cluster is in virial equilibrium, then $kT = G M^2/6Nr_{\rm vir}$ where $M$ is the total mass of the cluster and $r_{\rm vir}$ is the virial radius. Consequently, the binding energy of a binary with semi-major axis $a$ can be related to the descriptive parameters of the cluster by:
\begin{equation}
E_{\rm b} = 3 N \left(\frac{m_{\rm bh}}{M}\right)^2\frac{r_{\rm vir}}{a} kT.
\label{bulk_properties}
\end{equation}
Combining Equations~\ref{minimum_binding} and~\ref{bulk_properties} with Kepler's third law gives
\begin{equation}
P_{\rm min} = \left(\frac{r_{\rm vir}}{10 W_0 M}\right)^{3/2} \frac{\pi m_{\rm bh}}{\sqrt{G}}
\label{minimum_period}
\end{equation}
for the minimum orbital period of any remaining black hole binary in a globular cluster, where we have assumed that $N\langle{m}\rangle = M$.

We can estimate the range of periods for the remaining black hole binaries by inserting standard values for cluster parameters into Equation~\ref{minimum_period}. Using estimates from Takahashi and Portegies Zwart~\cite{takahashi00} of the parameters of zero-age globular clusters, we use $\log{r_{\rm vir}} = 0 \pm 0.3 {\rm pc}$, $\log{M} = 6.0 \pm 0.5 M_{\odot}$, and $W_0 = 7.5 \pm 2.5$ to determine this range. We find that the $P_{\rm orb} \sim 900 - 120,000 {\rm s}$ with the average value of $P_{\rm orb} = 4,200 {\rm s}$ for black hole masses of $m_{\rm bh} = 10 M_{\odot}$. Although this range is for the minimum orbital period of the remaining black hole binaries, it is not unreasonable to assume that the actual orbital periods will be close to the minimum value.

\section{Signal at LISA}

In a coordinate system centered on an eccentric binary with orbital frequency $f$, the gravitational wave at a distant detector located at angular position $\vartheta$ and $\varphi$ is given by the two polarizations~\cite{pierro01}:
\begin{eqnarray}
h_{\times} &=& \frac{\cos{\vartheta}}{\sqrt{2}}\sum_{n=1}^{\infty}{\left[2h_{xy}^{(n)}\sin{(2\pi nft)}\cos{2\varphi}-h_{x-y}^{(n)}\cos{(2\pi nft)}\sin{2\varphi}\right]}\label{hcross}\\
h_+ &=& \frac{1}{2\sqrt{2}}\sum_{n=1}^{\infty}{{\Big [}(1+\cos^2{\vartheta}){\Big (}2h_{xy}^{(n)}\sin{(2\pi nft)}\sin{2\varphi}}\nonumber\\
& & ~~~~~~~~~~~~~~~~~~~~~~~~~~~~~~~~~+h_{x-y}^{(n)}\cos{(2\pi nft)}\cos{2\varphi}{\Big )}\nonumber\\
& & ~~~~~~~~~~~~-(1-\cos^2{\vartheta})h_{x+y}^{(n)}\cos{(2\pi nft)}{\Big ]}\label{hplus},
\end{eqnarray}
where the metric components $h^{(n)}_{xy}$ and $h^{(n)}_{x\pm y}$ are:
\begin{eqnarray}
h^{(n)}_{xy} & = & h_0 n \left(1 - e^2\right)^{1/2}\left[J_{n-2}(ne) - 2J_{n}(ne) + J_{n+2}(ne)\right]\\
h^{(n)}_{x-y} & = & 2h_0 n {\Big \{}J_{n-2}(ne) - 2 e J_{n-1}(ne) + \frac{2}{n} J_n(ne)\nonumber\\
& & ~~~~~ + 2 e J_{n+1}(ne) - J_{n+2}(ne){\Big\}}\\
h^{(n)}_{x+y} & = & - 4 h_0 J_n(ne).
\end{eqnarray}
The common amplitude factor $h_0$ is:
\begin{equation}
h_0 = \frac{2G^{5/3}}{c^4}(2\pi f)^{2/3} {\cal M}^{5/3}
\label{h_nought}
\end{equation}
with the ``chirp mass'' ${\cal M}^{5/3} = M_1M_2(M_1+M_2)^{-1/3}$.

In the frequency domain, such a signal will appear in LISA as a series of monochromatic signals, each of which is modulated by the motion of the detector about the Sun. We can describe the signal received by LISA in terms of several defined unit vectors. Let the orientation of the arms of the constellation of the LISA spacecraft be given by the three vectors, $\mathbold{\ell}_1,~\mathbold{\ell}_2,~\mathbold{\ell}_3$. The direction from the sun to the source is $\hat{\mathbold{n}}$. The orientation of the binary orbit defined by $\hat{\mathbold{L}}$ (which points along the angular momentum vector), $\hat{\mathbold{a}}$ (which points along the semi-major axis toward periastron), and $\hat{\mathbold{b}} = \hat{\mathbold{L}} \times \hat{\mathbold{a}}$. The gravitational wave propagates along $-\hat{\mathbold{n}}$, and we define the polarization axes $\hat{\mathbold{p}} = \left(\hat{\mathbold{n}}\times\hat{\mathbold{L}}\right)/ \vert{\hat{\mathbold{n}}\times\hat{\mathbold{L}}}\vert$ and $\hat{\mathbold{q}} = - \hat{\mathbold{n}} \times \hat{\mathbold{p}}$. The angles used in Equations~\ref{hcross} and~\ref{hplus} are then defined by $\cos{\vartheta} = -\hat{\mathbold{n}}\cdot\hat{\mathbold{L}}$ and $\tan{\varphi} = \left(\hat{\mathbold{n}} \cdot \hat{\mathbold{b}}\right)/ \left(\hat{\mathbold{n}} \cdot \hat{\mathbold{a}}\right)$. The sensitivity of one interferometer of LISA to both polarizations can then be described by $F^+$ and $F^{\times}$, with:
\begin{eqnarray}
F^+ & = & (p_ap_b-q_aq_b)(\ell^a_1\ell^b_1 - \ell^a_2\ell^b_2)\\
F^{\times} & = & (p_aq_b-q_ap_b)(\ell^a_1\ell^b_1 - \ell^a_2\ell^b_2).
\label{antenna_pattern}
\end{eqnarray}
The signal received at LISA is then:
\begin{equation}
h(t) = \sum_{n=1}^{\infty}{\sqrt{A^2_n + B^2_n}\cos{\left[2\pi n f t + \phi_{np}(t) + \phi_{nD}(t)\right]}}
\label{signal}
\end{equation}
with
\begin{eqnarray}
A_n & = & \frac{1}{\sqrt{2}}\left(2F^{\times}\cos{\vartheta}\cos{2\varphi} + F^+\left(1+\cos^2{\vartheta}\right)\sin{2\varphi}\right)h^{(n)}_{xy}\\
B_n & = & \frac{-1}{2\sqrt{2}}{\Big [}\left(2F^{\times}\cos{\vartheta}\sin{2\varphi} - F^+\left(1 + \cos^2{\vartheta}\right)\cos{2\varphi}\right)h^{(n)}_{x-y}\nonumber\\
& & ~~~~~~~~~~+ F^+\left(1 - \cos^2{\vartheta}\right)h^{(n)}_{x+y}{\Big ]}\\
\phi_{np} & = & \tan^{-1}\left(-\frac{A_n}{B_n}\right)\\
\phi_{nD} & = & 2\pi n f \left(\frac{R}{c}\right) \sin{\theta_s}\cos{\left(\phi(t)-\phi_s\right)}.
\label{definitions}
\end{eqnarray}
The angular position of the source in heliocentric coordinates is $theta_s$ and $\phi_s$. The center-of-mass trajectory of LISA is $\phi(t) = 2\pi t/T$ with $T = 1~{\rm yr}$ and $R = 1~{\rm AU}$. A second signal can be synthesized from a linear combination of the responses from all three arms of LISA. This signal is equivalent to the signal from a detector rotated by $45^{\circ}$ in the plane of the initial detector~\cite{cutler98}.

\begin{figure}
\begin{center}
\def\epsfsize#1#2{0.7#1}
\epsfbox{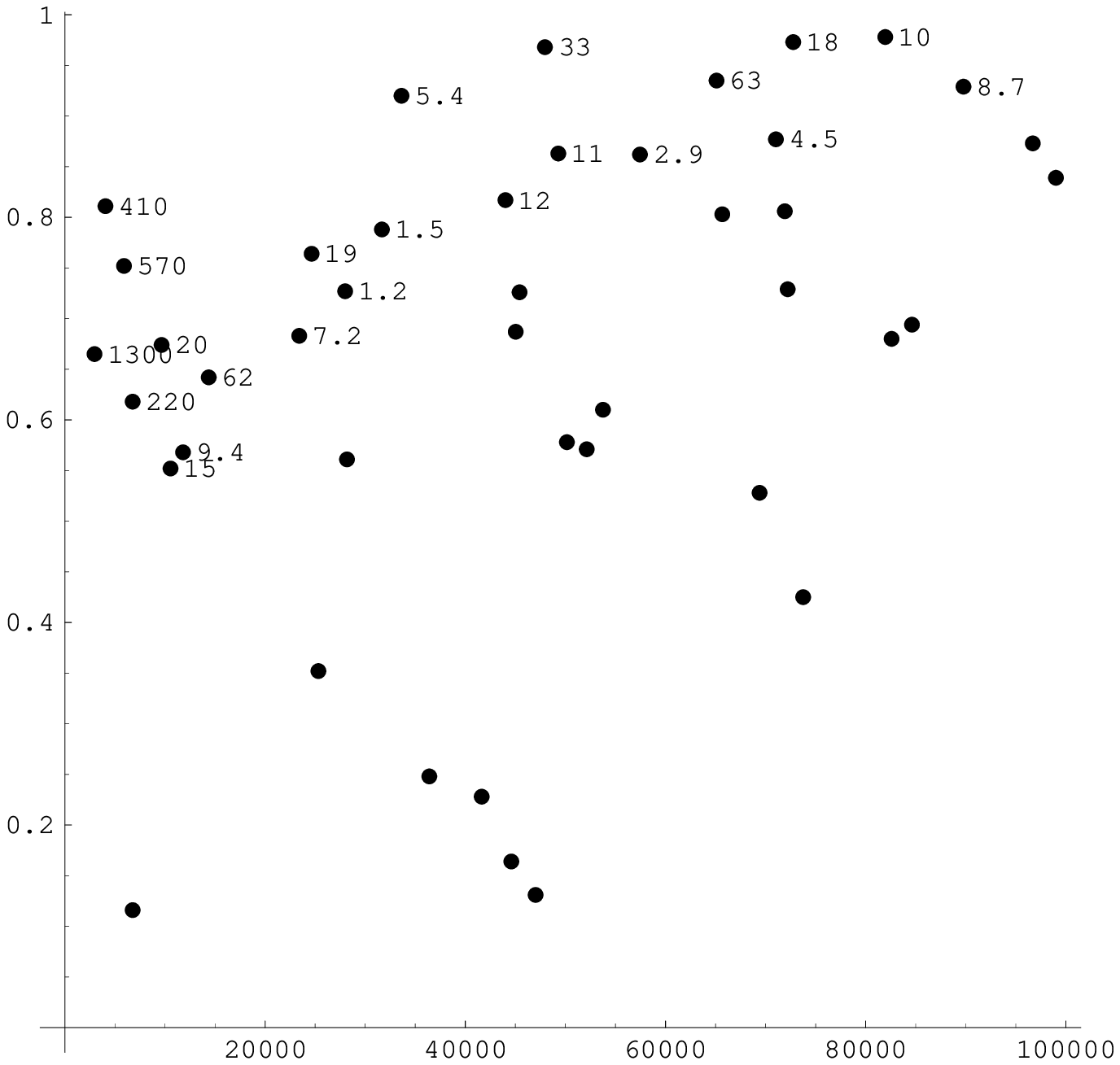}
\caption{Signal-to-noise values for all black hole binaries in the simulation. The data points are plotted for orbital period (in seconds) and eccentricity.}
\end{center}
\label{signal_strength}
\end{figure}

We have performed a study of the gravitational wave signal from a Monte Carlo simulation of the population of black hole binaries in globular clusters. We have chosen a pessimistic retention probability of $1/3$, so that there are 43 black hole binaries in the galactic globular cluster system. The black hole masses were chosen randomly from the range $6 - 18 M_{\odot}$. The eccentricities were chosen from a thermal distribution $(P(e) = 2e$, and the orbital period was selected from a flat distribution in the range $2,000 - 100,000 {\rm s}$. The angular resolution of the gravitational wave signals from these black holes was determined by looking at the first 10 harmonics above 1 mHz. These black hole binaries were assigned random globular cluster hosts, and the signal-to-noise was then calculated using the expected instrument noise and transfer function for LISA calculated by Larson and Hiscock~\cite{larson00}. The results show that fourteen of these had signal-to-noise greater than 10. Figure~\ref{signal_strength} shows the eccentricities and orbital periods for all 43 black hole binaries in the simulation.

\begin{figure}
\begin{center}
\def\epsfsize#1#2{0.7#1}
\epsfbox{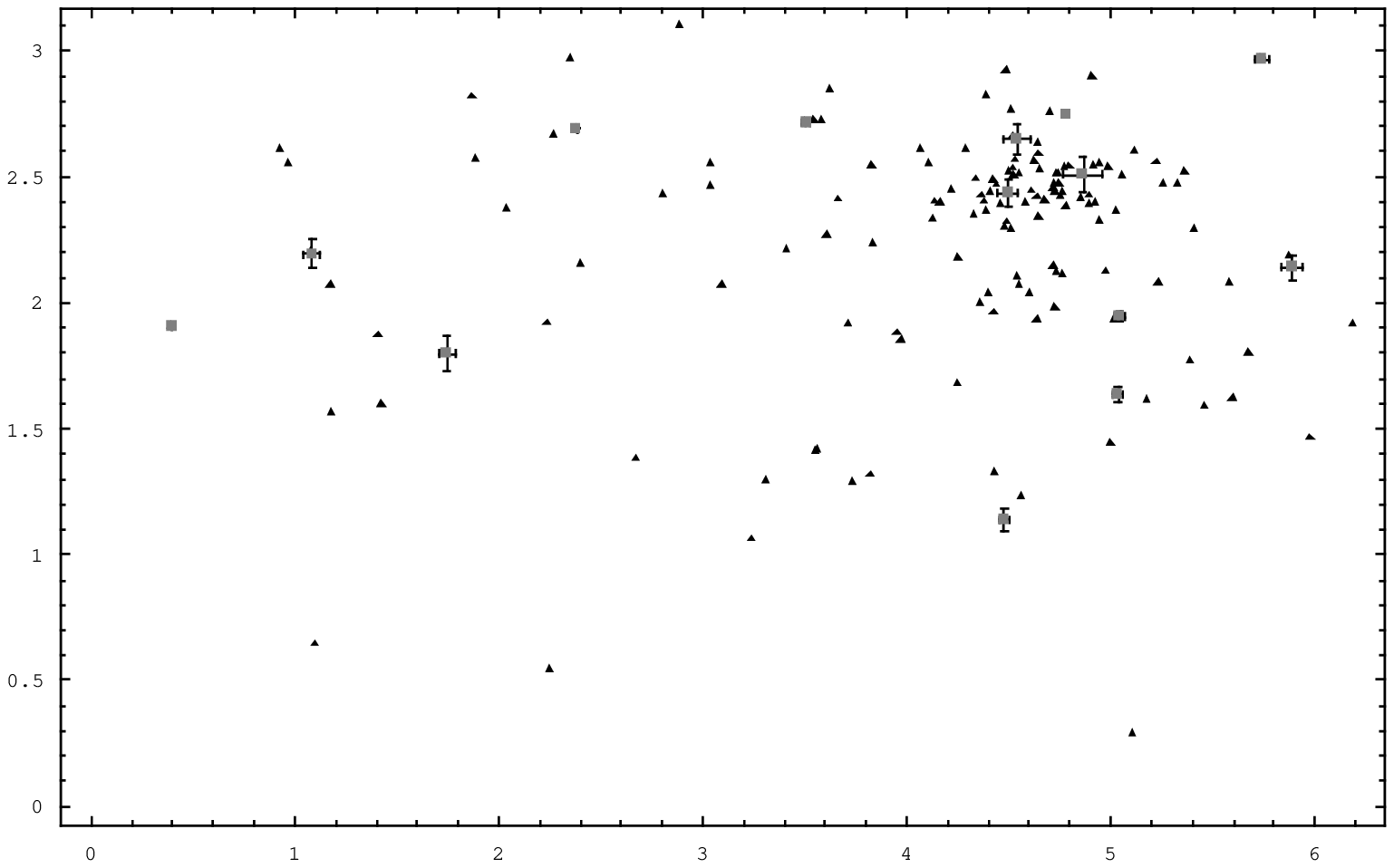}
\caption{Galactic globular cluster system in heliocentric coordinates. Black hole binaries are shown as gray squares, while the globular clusters are black triangles. The angular resolution for black hole binaries with error boxes small enough to allow for identification of the host cluster is shown.}
\end{center}
\label{angular_resolution}
\end{figure}

The Fisher matrix formalism to find the expected variances for the parameterization of the signal from eccentric binaries allows us to find the angular resolution for the binaries in the simulation. Using only the first five harmonics above 1 mHz, we find that the fourteen binaries with signal-to-noise above 10 also have sufficient angular resolution to identify them with their host clusters. The location and angular resolution for these 14 binaries are shown with the galactic globular cluster system in Figure~\ref{angular_resolution}.

\section{Conclusions}

The galactic globular cluster system may contain numerous black hole binaries. The evolution of stars in the early life of a globular cluster nearly guarantee that there will be several stellar mass black holes in each globular cluster. Through dynamical interactions and mass segregation, these black holes are expected to form a separate population within the cores of globular clusters. Further dynamical interactions between the black holes  will generate a population of black hole binaries which will subsequently be ejected from the globular cluster through recoil. However, it is expected that the final three black holes in the cluster will either eject each other, or result in a remaining binary system. During the interaction, the probability of retention can be estimated to be somewhere between $1/3$ and $1/2$. These retained binaries will have orbital frequencies below the gravitational wave frequency band of LISA. However, they are also expected to have large eccentricities. The luminosity in gravitational radiation from these binaries will be spread out among many of the higher harmonics, and consequently into LISA frequency band. We have performed a Monte Carlo simulation showing that $\sim 10\%$ of the globular clusters may contain a black hole binary with sufficient signal strength in the higher harmonics to allow for the angular position of the binary to be resolved to within the host cluster. Because the period distribution and likelihood of ejection for the black hole binaries are related to the structure and past history of the cluster, the identification and study of these black hole binaries may provide information about the past structure and history of their host clusters.

\ack
This work was supported in part by NASA grant NCC5-579, NASA EPSCoR grant NCCW-0058, and Montana EPSCoR grant NCC5-240. The author also acknowledges numerous useful discussions with Simon Portegies Zwart and Steve McMillan while attending the Workshop on Compact Objects in Dense Star Clusters at the Aspen Center for Physics in June, 2001.

\end{document}